\documentclass[preprint,floatfix] {revtex4} 
\newcommand{\rvec}{\mathrm {\mathbf {r}}} 
\newcommand{\pvec}{\mathrm {\mathbf {p}}} 
\usepackage{graphicx}
\usepackage{subfigure}
\usepackage{xcolor}
\usepackage{amsmath}
\usepackage{enumerate}
\usepackage{tabularx}
\usepackage{multirow}
\usepackage{color, soul}
\definecolor{darkblue}{rgb}{0,0,0.5}
\setulcolor{darkblue}
\renewcommand{\arraystretch}{1.0}
\begin{document}

\title{Relative Fisher information in some central potentials}

\author{Neetik Mukherjee}
%%\altaffiliation{Email: neetik.mukherjee@iiserkol.ac.in.}

\author{Amlan K.~Roy}
\altaffiliation{Corresponding author. Email: akroy@iiserkol.ac.in, akroy6k@gmail.com.}
\affiliation{Department of Chemical Sciences\\
Indian Institute of Science Education and Research (IISER) Kolkata, 
Mohanpur-741246, Nadia, WB, India}

\begin{abstract}
%%1234567890 %%1234567890 %%1234567890 %%1234567890 %%1234567890 %%1234567890 %%1234567890 %%1234567890 %%1234567890 %%1234567890
Relative Fisher information (IR), which is a measure of correlative fluctuation between two probability densities, has been 
pursued for a number of quantum systems, such as, 1D quantum harmonic oscillator (QHO) and a few central potentials namely, 
3D isotropic QHO, hydrogen atom and pseudoharmonic potential (PHP) in both position ($r$) and momentum ($p$) spaces. In the 1D 
case, the $n=0$ state is chosen as reference, whereas for a central potential, the respective circular or node-less 
(corresponding to lowest radial quantum number $n_{r}$) state of a given $l$ quantum number, is selected. 
Starting from their exact wave functions, expressions of IR in both $r$ and $p$ spaces are obtained in closed analytical 
forms in all these systems. A careful analysis reveals that, for the 1D QHO, IR in both coordinate spaces increase linearly 
with quantum number $n$. Likewise, for 3D QHO and PHP, it varies with single power of radial quantum number 
$n_{r}$ in both spaces. But, in H atom they depend on both principal ($n$) and azimuthal ($l$) quantum numbers. However, at a 
fixed $l$, IR (in conjugate spaces) initially advance with rise of $n$ and then falls off; also for a given $n$, it always 
decreases with $l$.

\vspace{3mm}


\vspace{3mm}
\end{abstract}
\maketitle

\section{introduction}
Over the years information theoretical concepts have emerged as certain valuable tool for analyzing various physical and 
chemical systems. Since inception, they have provided major impetus in many diverse fields of science and technology 
\cite{sen12}. These measures quantify the spatial distribution of single-particle density of a system in different complementary 
ways. Arguably, these are the most appropriate uncertainty measures, as they do not make any reference to some specific point 
of the corresponding Hilbert space. Moreover, these are closely related to energetics and experimentally measurable quantities 
of a system. R\'enyi entropy (R) is considered as information-generating functional, being directly connected to entropic 
moments, and completely characterizes density. Shannon entropy (S) is a special case of R, which provides the expectation 
values of logarithmic probability density. On the other hand, Fisher information (I) is a gradient functional of density 
that determines the fluctuation in a density distribution. In recent times, S has been extensively utilized to handle divergent 
perturbation series \cite{bender87}, for image reconstruction and spectral interpretation \cite{skilling91}, in polymer science 
\cite{penna03}, thermodynamics \cite{poland00,singer04,antoniazzi07} etc. The idea of maximization of information entropy with 
known values of first few moments was exploited exhaustively for a wide range of problems to search ground quantum stationary 
states \cite{plastino95,canosa89}. Lately they were used to investigate the competing behavior of localization-delocalization 
in double-well potential \cite{mukherjee15, mukherjee16}, effect of trapping of hydrogen atom in a spherically 
confined environment \cite{mukherjee18}. Applications of S, I are also found in formulation of Euler equation in orbital-free 
density functional theory \cite{nagy15}. Because of their ability to predict and explain versatile features of a system, 
they were invoked to explore a multitude of phenomena such as Pauli effects \cite{toranzo14,nagy06a}, ionization potential, 
polarizability \cite{sen07}, entanglement \cite{nagy06b}, avoided crossing \cite{ferez05} and so forth in atoms. In 
molecules, these quantities are used to explain steric effect \cite{nagy07,esquivel11}, bond formations 
\cite{nalewajski08}, elementary chemical reactions \cite{rosa10} etc., to cite a few.

In recent years, much attention was paid to examine several information measures like R,~S,~I, Tsallis entropy (T), Onicescu 
energy (E) in central potentials having relevance in physical and chemical problems. In this scenario, I in $r$ and 
$p$ spaces are expressible in terms of four radial expectation values, \emph{viz.}, $\langle p^{2} \rangle,~\langle r^{-2} 
\rangle$ and $\langle r^{2} \rangle,~\langle p^{-2} \rangle$ respectively \cite{romera05}. The product of these two quantities 
is bounded by both upper and 
lower limits given as, $\frac{81}{\langle r^2\rangle \langle p^2\rangle} \leq I_{\rvec} I_{\pvec} \leq 16 \ \langle r^2\rangle 
\langle p^2\rangle$ \cite{romera05}. A detailed inspection of R,~S,~I was recorded numerically in case of molecular PHP 
(for five diatomic molecules, Na$_{2}$, Cl$_{2}$, O$_{2}^{+}$, N$_{2}^{+}$, NO$^{+}$) \cite{yahya15}. 
Analogous calculation of S, I in composite $r$, $p$ spaces were reported for P\"oschl-Teller \cite{sun13}, Rosen-Morse 
\cite{sun13a}, squared tangent well \cite{dong14}, hyperbolic \cite{valencia15}, position-dependent mass Schr\"odinger equation 
\cite{chinphysb,yanez14} infinite circular well\cite{song15}, hyperbolic double-well \cite{sun15} potential, etc. The 
literature is quite vast and we have cited only a few selective ones. 

Kullback-Leibler divergence or relative entropy is an indicator of how one probability distribution function shifts from a 
given distribution \cite{kullback51,kullback78}. In quantum mechanics, this characterizes a measure of distinguishability 
between two states. It actually extracts the change of information from one state to other \cite{frieden04}. Relative R 
and S was studied for various atomic systems using H atom ground state as reference \cite{sagar08}. A detailed exploration reveals  
that, they are directly connected with atomic radii and quantum capacitance \cite{nagy09,sagar08}. Another 
interesting measure which has gained considerable popularity in past few years is the so-called relative Fisher information 
(IR) \cite{villani00}. It has distinct role in different topics of physics and chemistry, such as to calculate phase-space 
gradient of dissipated work and information \cite{yamano13}, deriving Jensen divergence \cite{sanchez12}, relation with score 
function \cite{toscani17}, in the context of study of probability current \cite{yamano13a}, in thermodynamics \cite{friden99}, 
etc. Very recently it has been profitably used in deriving atomic densities \cite{antolin09} and formulating density 
functionals under local-density and generalized-gradient approximations \cite{levamaki17}. Further, IR along with 
Hellmann-Feynman, and virial theorem has uncovered a Legendre transform structure related to Schr\"odinger equation 
\cite{flego11}. It has been derived self-consistently on the basis of estimation theory \cite{friden10}. In quantum chemistry 
perspective, it has been successfully derived using above two theorems and entropy maximization principle \cite{flego11a,
venkatesan14,venkatesan15}. Of late, radial IR for H atom in $r$ space has been estimated \cite{yamano18} numerically using 
ground state ($1s$) as reference. It would be nice and desirable to analyze this in $p$ space for H atom, as well as other 
quantum systems of topical interest. In this 
communication our primary aim is to examine IR in both $r, p$ spaces for certain central potentials by picking the \emph{lowest 
state corresponding to a given $l$}, as the reference for respective $l$-state calculations. Note that in literature, usually 
\emph{the lowest state} is chosen as reference. Elucidative calculations are performed for three important potentials, 
\emph{viz.}, 3D QHO, H atom and PHP. Apart from the work in \cite{yamano18} for IR$_{\rvec}$ (as mentioned above), such studies 
in central potentials are very scarce; more so in $p$ space, for which no work exists as yet, to the best of our knowledge. The 
current study derives simple analytical results for IR in two spaces, in all three central potentials considered here. 
Before doing that, we consider the prototypical case of a 1D QHO. In all cases, we take the exact analytical solution, to start 
with (in both spaces). For PHP case, the illustration is done by choosing six representative diatomic molecules (H$_2$, CO, O$_2^+$, 
Na$_2$, Cl$_2$, NO) including one cation. Numerical results are offered as appropriate, and comparison with literature works are
made, wherever possible. The organization of our article is as follows. Section II gives the essential details 
of our present formulation; then Sec.~III offers a discussion on the results, while we conclude with a few remarks in Sec.~IV.    

\section{Formulation} 
For two normalized probability densities $\rho_{n,l,m}(\tau)$ and $\rho_{n_{1},l_{1},m_{1}}(\tau)$, IR is expressed as,
\begin{equation}
\mathrm{IR} \ [\rho_{n,l,m}(\tau)|\rho_{n_{1},l_{1},m_{1}}(\tau)]=\int_{{\mathcal{R}}^3}\rho_{n,l,m}(\tau)\left|\nabla \ \mathrm{ln}
\left\{ \frac{\rho_{n,l,m}(\tau)}{\rho_{n_{1},l_{1},m_{1}} (\tau)}\right \} \right|^{2}  \mathrm{d}\tau.
\end{equation}
Here $n,l,m$ and $n_{1},l_{1},m_{1}$ are the identifiers of target and reference states respectively, while $\tau$ is a 
generalized variable. In case of central potential, these probability densities $\rho_{n,l,m}(\tau)$, $\rho_{n_{1},l_{1},m_{1}}
(\tau)$ can be written in following forms, without any loss of generality, 
\begin{equation}
\begin{aligned}
\rho_{n,l,m}(\tau) & = R_{n,l}^{2}(s) \Theta_{l,m}^{2}(\theta);  \      \hspace{30mm} 
\rho_{n_{1},l_{1},m_{1}}(\tau) = R_{n_{1},l_{1}}^{2}(s) \Theta_{l_{1},m_{1}}^{2}(\theta). 
\end{aligned}
\end{equation}
In the above equation, $R_{n,l}(s), R_{n_{1},l_{1}}(s)$ represent radial parts, $\Theta_{n,l}(\theta), \Theta_{n_{1},l_{1}}
(\theta)$ signify angular contributions of two wave functions, whereas $``s"$ implies either $r$ or $p$ variable in respective 
radial functions. Thus Eq.~(1) may be rewritten as, 
\begin{multline}
\mathrm{IR} \ [\rho_{n,l,m}(\tau)|\rho_{n_{1},l_{1},m_{1}}(\tau)]= \mathrm{IR} \ [\rho_{n,l}(s)|\rho_{n_{1},l_{1}}(s)]+
\left\langle \frac{1}{s^{2}} \right\rangle \mathrm{IR} \ [\Theta_{l,m}^{2}(\theta)|\Theta_{l_{1},m_{1}}^{2}(\theta)]+\\
2\int_{0}^{\infty}s~R_{n,l}^{2}(s)\left[\frac{d}{ds}\mathrm{ln}\left\{ \frac{R^{2}_{n,l}(s)}{R^{2}_{n_{1},l_{1}}(s)}\right\}
\right]\mathrm{d}s
\int_{0}^{\pi}\Theta_{l,m}^{2}(\theta)\left[\frac{d}{d\theta}\mathrm{ln}\left\{ \frac{\Theta_{l,m}^{2}(\theta)}
{\Theta_{l_{1},m_{1}}^{2}(\theta)}\right\} \right] \sin\theta \mathrm{d}\theta
\end{multline}
where the following quantities have been defined, 
\begin{equation}
\begin{aligned}
\mathrm{IR} \ [\rho_{n,l}(s)|\rho_{n_{1},l_{1}}(s)]& =\int_{0}^{\infty}s^{2}~\rho_{n,l}(s)\left|\frac{d}{ds}\mathrm{ln}
\left\{ \frac{\rho_{n,l}(s)}{\rho_{n_{1},l_{1}}(s)}\right\} \right|^{2}\mathrm{d}s \\ 
\mathrm{IR} \ [\Theta_{l,m}^{2}(\theta)|\Theta_{l_{1},m_{1}}^{2}(\theta)]&=\int_{0}^{\pi}\Theta_{l,m}^{2}(\theta)\left|
\frac{d}{d\theta}\mathrm{ln}\left\{ \frac{\Theta_{l,m}^{2}(\theta)}{\Theta_{l_{1},m_{1}}^{2}(\theta)}\right\} \right|^{2} 
\sin\theta~\mathrm{d}\theta      
\end{aligned}
\end{equation}
If one opts for $l=l_{1}$ and $m=m_{1}$, then angular portions of $\rho_{n,l,m}(\tau)$, $\rho_{n_{1},l_{1},m_{1}}(\tau)$ become identical. Under such 
condition, Eq.~(3) takes the following simplified form, 
\begin{equation}
\mathrm{IR} \ [\rho_{n,l,m}(s)|\rho_{n_{1},l,m}(s)]=\int_{0}^{\infty}s^{2}R_{n,l}^{2}(s)\left|\frac{d}{ds} \ 
\mathrm{ln}\left\{ \frac{R_{n,l}^{2}(s)}{R_{n_{1},l}^{2} (s)}\right\} \right|^{2}  \mathrm{d}s
\end{equation}
Clearly, the right-hand side of Eq.~(5) reduces to zero when the two radial probability densities are identical ($n=n_{1}$). 
Throughout the article, 
$R_{n_{1},l}(s)$ refers to the reference state and for the purposes of IR calculation, as such, it should always be a node-less 
distribution of $s$; otherwise IR will blow up. For a particular $l$, this study has considered the relevant circular state 
corresponding to that $l$ as standard, in composite $r,~p$ spaces. This choice of $l=l_1$ serves the basic purpose of dealing with 
the radial IR. For a H atom, it is also physically consistent, because the shape of an orbital exclusively depends on the value 
of $l$. Importantly, the distribution of an electron is determined by the shape of the orbital where it resides. Hence, it is 
sensible to compare a density-dependent property between the orbitals with resembling shape. Furthermore, for simplicity's sake, 
we also set $m=m_{1}$, which leads to cancellation of angular portion in logarithmic part of the integrand. Further, in full 
IR calculation, angular part normalizes to unity. Hence, IR remains invariant to magnetic quantum number. 

For the systems considered in this work, Eq.~(5) may further be recast to,  
\begin{equation}
\mathrm{IR}_s= 4\int_{0}^{\infty}s^{2}R_{n,l}^{2}(s)\left(\frac{R^{\prime}_{n,l}(s)}{R_{n,l}(s)}-\frac{R^{\prime}_{n_{1},l}(s)}
{R_{n_{1},l}(s)}\right)^{2}\mathrm{d}s.
\end{equation}
Now let us assume that, $\psi_{n,l}(s)=f_{l}(s)P_{n,l}(s)$, 
where $f_{l}(s)$ is such that its functional form depends on $l$ only. Here $P_{n,l}(s)$ is a polynomial of $s$, and prime 
denotes first derivative with respect to $s$. Hence one obtains,
\begin{equation}
\frac{R^{\prime}_{n,l}(s)}{R_{n,l}(s)}=\frac{f^{\prime}_{l}(s)}{f_{l}(s)}-\frac{P^{\prime}_{n,l}(s)}{P_{n,l}(s)}. 
\end{equation}
For a fixed $l$, $f_{l}(s)$ has identical mathematical form in reference and standard state; only the polynomial part changes with 
$n$. Therefore the target and reference state having same $l$ will only differ in the polynomial part, giving rise to, 
\begin{equation}
\frac{R^{\prime}_{n_{1},l}(s)}{R_{n_{1},l}(s)}=\frac{f^{\prime}_{l}(s)}{f_{l}(s)}-
\frac{P^{\prime}_{n_{1},l}(s)}{P_{n_{1},l}(s)}.
\end{equation}    
If node-less reference state is adopted, the ratio of the polynomial in right-hand side reduces to unity. Then Eq.~(8) may be 
rewritten as below, 
\begin{equation}
\frac{R^{\prime}_{n_{1},l}(s)}{R_{n_{1},l}(s)}=\frac{f^{\prime}_{l}(s)}{f_{l}(s)}. 
\end{equation}
Use of Eqs.~(7) and (9), results in,  
\begin{equation}
\frac{R^{\prime}_{n_{1},l}(s)}{R_{n_{1},l}(s)}-\frac{R^{\prime}_{n_{1},l}(s)}{R_{n_{1},l}(s)} = 
\frac{P^{\prime}_{n,l}(s)}{P_{n,l}(s)}. 
\end{equation}
Substitution of Eq.~(10) in Eq.~(6), finally  yields the following generalized form of  
\begin{equation}
\mathrm{IR}=4\int_{0}^{\infty}s^{2}f_{l}^{2}(s)\left[P_{n,l}^{\prime}(s)\right]^{2}\mathrm{d}s, 
\end{equation}  
in a central potential (when the reference and target states have same $l$).
 
\section{Result and Discussion}
\subsection{1D QHO}
Before proceeding for central potential, at first we would like to explore IR in a model 1D QHO in both $x, p$ spaces. 
The underlying potential is characterized by the expression: $v(x)= \frac{1}{2} \omega^2 x^2$ (mass $m$ is set to unity
throughout), where $\omega$ signifies angular frequency. The normalized $x$-space wave function is expressed as ($H_n (x)$ refers
to Hermite polynomial),
\begin{equation}
\psi_{n}(x)=\left(\frac{\omega}{\sqrt{2}\pi}\right)^\frac{1}{4}\frac{1}{\sqrt{2^{n}n!}} \ H_{n}\left(\frac{\sqrt{\omega}}
{2^{\frac{1}{4}}}x\right)e^{-\frac{\omega}{2\sqrt{2}}x^{2}}.
\end{equation}
Choosing $n=0$ as reference state, $\frac{\sqrt{\omega}}{2^{\frac{1}{4}}}x=y$, and using definition of IR in Eq.~(1), 
one gets, 
\begin{equation}
\mathrm{IR}_{x}=\frac{\omega}{\sqrt{2}} \ \frac{1}{2^{n-1}n!}\int_{0}^{\infty} \ H_{n}^{2}(y) \ 
e^{-y^{2}}\left[\frac{\psi^{\prime}_{n}(y)}
{\psi_{n}(y)}-\frac{\psi^{\prime}_{0}(y)}{\psi_{0}(y)}\right]^{2}\mathrm{d}y. \setlength{\arrayrulewidth}{1mm}
\setlength{\tabcolsep}{18pt}
\renewcommand{\arraystretch}{1.5}
\end{equation}
Here the suffix ``x" denotes a position-space quantity. 
Now use of recurrence relation $H_{n}^{\prime}(y)=2nH_{n-1}(y)$, in conjunction with orthonormality condition of Hermite 
polynomial, $\int_{0}^{\infty}e^{-y^{2}}H_{m}(y)H_{k}(y) \mathrm{d}y=2^{m-1}m!\sqrt{\pi}\delta_{mk}$ 
($\delta_{mn}=1$ when $m=k$, and $0$ otherwise) produces,  
\begin{equation}
\mathrm{IR}_{x} =4n^{2}\frac{\omega}{\sqrt{2}} \ \frac{1}{2^{n-1}n!}\int_{0}^{\infty}\left[H_{n-1}(y)\right]^{2}
e^{-y^{2}}\mathrm{d}y 
       =4\sqrt{2} \ \omega n. 
\end{equation}
Thus Eq.~(14) suggests that, IR$_{x}$ in $n$th state may be obtained from a knowledge of $(n-1)$th-state wave
function. Evidently, it increases linearly with state index, $n$, with a positive slope of $4\sqrt{2}\omega$. This
is in consonance with the fact that in this system, localization as well as fluctuation increase with $\omega$. So IR$_{x}$ 
result simply complements this. 

Now we move on to $p$ space, where the normalized wave function is given as, 
\begin{equation}
\psi_{n}(p)=\left(\frac{\sqrt{2}}{\omega \pi}\right)^\frac{1}{4}\frac{1}{\sqrt{2^{n}n!}} \ H_{n}\left(\frac{2^{\frac{1}{4}}}
{\sqrt{\omega}}p\right)e^{-\frac{1}{\sqrt{2}\omega}p^{2}}
\end{equation}   
Again, considering $n=0$ state as standard, setting $\frac{2^{\frac{1}{4}}}{\sqrt{\omega}}p=g$, using the recurrence relation 
$H_{n}^{\prime}(g)=2nH_{n-1}(g)$ and invoking orthonormality condition of Hermite polynomial (stated earlier), one can derive
the following expression for IR$_{p}$ after some straightforward algebra ($p$ subscript indicates $p$-space quantity), namely, 
\begin{equation}
\mathrm{IR}_{p}=4n^{2}\sqrt{\frac{2}{\pi}} \ \frac{1}{2^{n}n!}\frac{1}{\omega}\int_{0}^{\infty}[H_{n-1}(g)]^{2}e^{-g^{2}}\mathrm{d}g 
      =\frac{8\sqrt{2}}{\omega}n. 
\end{equation}
Thus, similar to IR$_{x}$, here also IR$_{p}$ of a given oscillator state can be recovered  from the wave function of 
adjacent lower state. Equation~(16) implies that, progress of IR$_{p}$ with $n$ is again linear like its $x$-space counterpart,
slope of the straight line in this case being $\frac{8\sqrt{2}}{\omega}$. It is inversely proportional to $\omega$ in accordance 
with the fact that, an increase in oscillation enhances localization as well as fluctuation. At the special value of 
$\omega=\sqrt{2}$, IR$_{x}$, IR$_{p}$ become equal ($8n$). Further, like the total energy difference in a QHO, 
$\Delta \mathrm{E_{n}}(=\mathrm{E_{n+1}}-\mathrm{E_{n}})$, the difference of IR between two successive states also remains
constant, i.e.,  $\Delta (\mathrm{IR}_{x})= \mathrm{IR}_{x}(n+1)-\mathrm{IR}_{x}(n)=4\sqrt{2}\omega$, and 
$\Delta (\mathrm{IR}_{p})=\mathrm{IR}_{p}(n+1)-\mathrm{IR}_{p}(n)=\frac{8\sqrt{2}}{\omega}$.

\subsection{Central potential}
This subsection is now devoted to IR in central potentials. The quantum-mechanical probability density of the bound state of a 
non-relativistic particle in such a potential is obtained from the corresponding wave function, which in turn is determined from 
the solution of pertinent Schr\"odinger equation ($n_r, l$ signify radial and azimuthal quantum numbers),
\begin{equation}
\left[-\frac{1}{2}\nabla^{2}+v(r)\right]\psi_{n_r,l,m} (\rvec)=\mathcal{E}_{n_r,l} \ \psi_{n_r,l,m}(\rvec). 
\end{equation}
Where $\mathcal{E}_{n_r,l}$ is the energy of the state represented by $\psi_{n_r,l,m}(\rvec)$. In what follows, atomic unit is 
used unless otherwise mentioned. The spherical symmetry of central potentials permits one to split the wave function into radial 
and angular segments in spherical polar coordinates. The radial eigenfunction $R_{n_r,l}(r)$ obeys the radial differential 
equation,
\begin{equation}
\left[-\frac{1}{2}\frac{d^{2}}{dr^{2}}-\frac{1}{r}\frac{d}{dr}+\frac{l(l+1)}{2r^{2}}+v(r)\right]R_{n_r,l}(r)=
\mathcal{E}_{n_r,l}R_{n_r,l}(r)
\end{equation}
Our interest lies in three important potentials corresponding to following functional forms for $v(r)$: (i) $\frac{1}{2} 
\omega^2 r^2$ ($\omega $ is oscillation frequency), the 3D QHO (ii) $-\frac{Z}{r}$ for H-like atom ($Z$ is atomic number) and 
(iii) $D_{e}\left(\frac{r}{r_{e}}-\frac{r_{e}}{r}\right)^{2}$, with $D_{e}, r_{e}$ representing dissociation energy and 
equilibrium intermolecular separation, in case of a PHP.

The $p$-space wave function for a particle in a central potential is obtained from respective Fourier transform of the $r$-space 
counterpart, and as such, is given below,
\begin{equation}
R_{n_r,l}(p)  =  \frac{1}{(2\pi)^{\frac{3}{2}}} \  \int_0^\infty \int_0^\pi \int_0^{2\pi} R_{n_r,l}(r) \ \Theta(\theta) 
 \Phi(\phi) \ e^{ipr \cos \theta}  r^2 \sin \theta \ \mathrm{d}r \mathrm{d} \theta \mathrm{d} \phi.
\end{equation}
Note that $R_{n_r,l}(p)$ is not normalized; thus needs to be normalized. If the radial functions are real, then Eq.~(6) assumes 
the following form,
\begin{equation}
\mathrm{IR} \ [\rho_{n_r,l}(s)|\rho_{n_{r_1},l}(s)]  = 4\int_{0}^{\infty}s^{2}R_{n_r,l}^{2}(s)\left[\frac{R_{n_r,l}^{\prime}(s)}
{R_{n_r,l}(s)}- \frac{R_{n_{r_1},l}^{\prime}(s)}{R_{n_{r_1},l}(s)}\right]^{2}  \mathrm{d}s.
\end{equation}
In the following, we attempt to derive IR$_{\rvec}$ and IR$_{\pvec}$ in the three prototypical systems mentioned earlier, starting 
from Eq.~(20).

\subsubsection{Isotropic 3D QHO}
We start from the normalized $r$-space wave function ($n_r$, the radial quantum number is related to $n$ as $n=2n_r+l$) given 
below, 
\begin{equation}
\psi_{n_{r},l}(r)= \sqrt{\frac{2 \ \omega^{l+\frac{3}{2}} \ n_{r}!}{\Gamma(n_{r}+l+\frac{3}{2})}} \ \ 
r^{l} \ e^{-\frac{\omega r^{2}}{2}} \ L_{n_{r}}^{l+\frac{1}{2}}(\omega r^{2}).
\end{equation} 
In the above, $L_n^{\alpha}(x)$ represents the associated Laguerre polynomial. Now, using Eq.~(21) and substituting
$\omega r^{2}=u$, Eq.~(20) yields,     
\begin{equation}
\mathrm{IR}_{\rvec}=\frac{16 \ \omega \ n_{r}!}{\Gamma(n_{r}+l+\frac{3}{2})} \int_{0}^{\infty} u^{l+\frac{3}{2}} e^{-u} 
\left[L_{n_{r}}^{l+\frac{1}{2}}(u)\right]^{2} \left(\frac{\psi_{n_{r},l}^{\prime}(u)}{\psi_{n_{r},l}(u)}-
\frac{\psi_{n_{r_{1}},l}^{\prime}(u)}{\psi_{n_{r_{1}},l}(u)}\right)^{2} \mathrm{d}u. 
\end{equation} 
Now, using the well-known recurrence relation $\ \frac{d}{du}L_{n_r}^{l+\frac{1}{2}} (u)=-L_{n_r-1}^{l+\frac{3}{2}} (u)$, the 
ratios of wave functions occurring in the parentheses may be simplified as, 
\begin{equation}
\frac{\psi_{n_{r},l}^{\prime}(u)}{\psi_{n_{r},l}(u)}= 
\frac{l}{2u}-\frac{1}{2}-\frac{L_{n_{r}-1}^{l+\frac{3}{2}} (u) }{L_{n_{r}}^{l+\frac{1}{2}} (u) }, 
\end{equation} 
the right-hand side of which, for a node-less state becomes $\left(\frac{l}{2u}-\frac{1}{2}\right)$, because the ratio of 
polynomials vanishes. Now one may invoke the familiar orthonormality relation, 
\begin{equation} 
\int_{0}^{\infty}u^{k} e^{-u} L_{i}^{k}(u) L_{j}^{k}(u) \mathrm{d}u=\frac{(i+k)!}{i!} \ \delta_{ij}. 
\end{equation}
the final form of IR$_{\rvec}$ turns out as below,
\begin{equation}
\mathrm{IR}_{\rvec}= \frac{16 \ \omega \ n_r!}{\Gamma(n_{r}+l+\frac{3}{2})} \  \int_{0}^{\infty} u^{l+\frac{3}{2}} e^{-u} 
\left[L_{n_{r}-1}^{l+\frac{3}{2}}\right]^{2} \mathrm{d}u =16\omega \ n_{r} = 8\omega (n-l).  
\end{equation} 
Equation~(25) predicts that, IR$_{\rvec}$ in a 3D QHO, like its 1D counterpart, is also a linear function of $n$; however in this
occasion the slope is $8\omega$ (in contrast to $4 \sqrt{2}\omega$) and intercept is less than zero (in contrast 
to zero in 1D). For a fixed $l$, the slope increases and intercept decreases with $\omega$ respectively. Further, at a certain 
$\omega$, the intercept falls off with rise of $l$. In this scenario, the particle gets more and more localized with growth of 
$\omega$. The relative fluctuation with respect to reference state increases with $\omega$. Dependence of IR$_{\rvec}$ on $\omega$ 
is reminiscent to that of $r$-space Fisher information, I$_{\rvec}$ \cite{romera05}--both quantities escalate with $\omega$. 

Analogously, in $p$-space the normalized wave function has the form \cite{yanez94},  
\begin{equation}
\psi_{n_{r},l}(p)=\sqrt{\frac{2 \ n_{r}!}{\Gamma(n_{r}+l+\frac{3}{2}) \ \ \omega^{l+\frac{3}{2}}}} \ \ 
p^{l} \ e^{-\frac{p^{2}}{2\omega}} \ L_{n_{r}}^{l+\frac{1}{2}}\left(\frac{ p^{2}}{\omega}\right)
\end{equation}
Substituting $\frac{p^{2}}{\omega}=\chi$, and going through some simple algebraic steps, one can derive, 
\begin{equation}
\mathrm{IR}_{\pvec}=\frac{16 \ n_{r}!}{\Gamma(n_{r}+l+\frac{3}{2}) \ \omega}\ \int_{0}^{\infty} 
\chi^{l+\frac{3}{2}} e^{-{\chi}} 
\left[L_{n_{r}}^{l+\frac{1}{2}}(\chi)\right]^{2} \left(\frac{\psi_{n_{r},l}^{\prime}(\chi)}{\psi_{n_{r},l}(\chi)}-
\frac{\psi_{n_{r_{1}},l}^{\prime}(\chi)}{\psi_{n_{r_{1}},l}(\chi)}\right)^{2} \mathrm{d}\chi.
\end{equation}
Applying similar arguments as discussed earlier for IR$_{\rvec}$ produces,  
\begin{equation}
\frac{\psi_{n_{r},l}^{\prime}(\chi)}{\psi_{n_{r},l}(\chi)}-\frac{\psi_{n_{r_{1}},l}^{\prime}(\chi)}{\psi_{n_{r_{1}},l}(\chi)}=
-\frac{L_{n_{r}-1}^{l+\frac{3}{2}} (\chi) }{L_{n_{r}}^{l+\frac{1}{2}} (\chi) }. 
\end{equation}
which upon applying in Eq.~(27), leads to the following final expression, namely, 
\begin{equation}
\mathrm{IR}_{\pvec}=16\left[\frac{n_{r}!}{\Gamma(n_{r}+l+\frac{3}{2}) \ \omega}\right]\int_{0}^{\infty} 
{\chi}^{l+\frac{3}{2}}e^{-\chi}\left[L_{n_{r}-1}^{l+\frac{3}{2}}\right]^{2} \mathrm{d} \chi
      =\frac{16}{\omega}n_{r} = \frac{8}{\omega}(n-l).
\end{equation}
Equation~(29) indicates that, IR$_{\pvec}$, like IR$_{\rvec}$, also linearly varies with $n$; the slope and intercept being
$\frac{16}{\omega}$ and $-\frac{16l}{\omega}$ respectively. An increase in $\omega$ facilitates localization and hence 
consequently fluctuation too. IR$_{\rvec}$ rises with $\omega$, while IR$_{\pvec}$ falls off, signifying higher 
fluctuation at larger $n_r$. Once again IR$_{\rvec}$, IR$_{\pvec}$ of a given $n_r,l$-state may be calculated from 
$(n_{r}-1),~(l+1)$-state wave functions; this holds true in both spaces. We close the discussion by noting that, in parallel 
to 1D case, here also both $\Delta (\mathrm{IR}_{\rvec})$, $\Delta (\mathrm{IR}_{\pvec})$, at a given $l$, depend only on $\omega$ 
and remain unchanged with respect to $n$. They are expressed as; 
\begin{equation}
\begin{aligned}
\Delta (\mathrm{IR}_{\rvec}) & = \mathrm{IR}_{\rvec}(n+1,~l)-\mathrm{IR}_{\rvec}(n,~l)=8\omega \\    
\Delta (\mathrm{IR}_{\pvec}) & = \mathrm{IR}_{\pvec}(n+1,~l)-\mathrm{IR}_{\pvec}(n,~l)=\frac{8}{\omega}. 
\end{aligned}
\end{equation}

\begingroup           %%Table 1
\squeezetable
\begin{table}
\caption{Some specimen IR$_{\rvec}$ results for $n_ss, n_pp, n_dd, n_ff \ (n_s \geq 2, n_p \geq 3, n_d \geq 4, n_f \geq 5)$
orbitals of H atom, considering the corresponding circular states as reference. For $2s, 3s, 4s, 5s$ 
states, data given in parentheses represent literature results \cite{yamano18}. See text for details.}
\begin{ruledtabular}
\begin{tabular}{cl|cl|cl|cl} 
Orbital   &  IR$_{\rvec}$                &  Orbital  &  IR$_{\rvec}$      &
Orbital   &  IR$_{\rvec}$                &  Orbital  &  IR$_{\rvec}$      \\  \hline
$2s$    & 1                (1)           &  $3p$   & $\frac{8}{27}$ & $4d$ &  $\frac{1}{8}$        &  $5f$ & $\frac{8}{125}$ \\
$3s$    & $\frac{16}{27}$  (0.5925)      &  $4p$   & $\frac{1}{4}$      & $5d$ &  $\frac{16}{125}$     &  $6f$ & $\frac{2}{27}$ \\
$4s$    & $\frac{3}{8}$    (0.375)       &  $5p$   & $\frac{24}{125}$    & $6d$ &  $\frac{1}{9}$     &  $7f$ & $\frac{24}{343}$ \\
$5s$    & $\frac{32}{125}$ (0.256)       &  $6p$   & $\frac{4}{27}$ & $7d$ &  $\frac{32}{343}$ &  $8f$ & $\frac{1}{6}$ \\
\end{tabular}
\end{ruledtabular}
\end{table}
\endgroup

%%start from here
\subsubsection{H-isoelectronic series}
Our starting point is the radial function in $r$ space ($n$ signify radial quantum number), 
\begin{equation}
\psi_{n,l}(r)= \frac{2}{n^2}\left[\frac{(n-l-1)!}{(n+l)!}\right]^{\frac{1}{2}}\left[\frac{2Z}{n}r\right]^{l} 
e^{-\frac{Z}{n}r} \ L_{(n-l-1)}^{(2l+1)} \left(\frac{2Z}{n}r\right).  
\end{equation}
Again $L_n^{\alpha}(x)$ has usual meaning. Putting $\xi=\frac{2Zr}{n}$ in Eq.~(20), one gets,
\begin{equation} 
\mathrm{IR}_{\rvec} = 4 \left(\frac{n}{2Z}\right) \int_{0}^{\infty} \xi^{2}R_{n,l}^{2}(\xi)\left[\frac{R_{n,l}^{\prime}(\xi)}
{R_{n,l}(\xi)}-\frac{R_{n_{1},l}^{\prime}(\xi)}{R_{n_{1},l}(\xi)}\right]^{2} \mathrm{d} \xi .
\end{equation}
Where primes denote 1st-order derivatives with respect to $\xi$. Then one can write, 
\begin{equation}
\frac{R_{n,l}^{\prime}(\xi)}{R_{n,l}(\xi)}=\frac{l}{\xi}-\frac{1}{2}-\frac{L_{(n-l-2)}^{(2l+2)}(\xi)}{L_{(n-l-1)}^{(2l+1)}(\xi)};  
\ \ \ \ \ \ \ \ \ \ \frac{R_{n_{1},l}^{\prime}(\xi)}{R_{n_{1},l}(\xi)}=\frac{l}{\xi}-\frac{1}{2}-
\frac{L_{(n_{1}-l-2)}^{(2l+2)}(\xi)} {L_{(n_{1}-l-1)}^{(2l+1)}(\xi)}.
\end{equation}
Since $R_{n_{1},l}(\xi)$ corresponds to a circular state, $L_{(n_{1}-l-1)}^{(2l+1)}(\xi)$ provides a 
constant term, and hence $L_{(n_{1}-l-2)}^{(2l+2)}(\xi)=0$. This simplifies the second ratio as, 
$\frac{R_{n_{1},l}^{\prime}(\xi)}{R_{n_{1},l}(\xi)}=\frac{l}{\xi}-\frac{1}{2}$. Using above condition and 
orthonormality condition of $L_n^{\alpha}(x)$, Eq.~(32) gives the following,  
\begin{equation} 
\mathrm{IR}_{\rvec} = \left(\frac{8}{Zn^3}\right)\frac{(n-l-1)!}{(n+l)!} \int_{0}^{\infty} \xi^{2l+2} e^{-\xi} 
\left[L_{(n-l-2)}^{(2l+2)}(\xi)\right]^{2} \mathrm{d} \xi,
\end{equation}    
which, after some algebraic manipulation, gives the final form of IR$_{\rvec}$ as below, 
\begin{equation}
\mathrm{IR}_{\rvec}=\frac{8 (n-l-1)}{Zn^3}  \ \ \ \ \ \ \ \  (\mathrm{when} \ \ n>l,  \ n-l \geq 2).
\end{equation}

Equation~(31) clearly indicates that, IR$_{\rvec}$ reduces with rise of $n,~l$ and $Z$. Thus with progress in $n,~l$, the spatial 
separation between two distributions deteriorates. In other words, the fluctuation of a particular state with respect to 
reference state reduces with the addition of nodes. It may be recalled that the behavioral pattern of IR$_{\rvec}$ with $n$ is 
akin to that of Fisher information in $r$ space, I$_{\rvec}$ \cite{romera05}--both decline as $n$
grows. However, I$_{\rvec}$ is invariant of $l$, whereas, IR$_{\rvec}$ seems to lessen with growth of $l$, for a certain $n$. 
Table~I offers some representative IR$_{\rvec}$ for $n_ss, n_pp, n_dd, n_ff (n_s \geq 2, n_p \geq 3, n_d \geq 4, n_f \geq 5$)
orbitals for H atom. For the $ns$ series, these have been published very recently (considering $1s$ as reference), which are 
duly quoted and compared with present work. As seen, the two results are practically identical. For the \emph{non-zero}-$l$ 
states however, we are not aware of any such reporting, and we offer here the first-time results on these. 

In case of \emph{even}-$l$ states maximum in IR$_{\rvec}$ appears at $n=\frac{(3l+4)}{2}$, the corresponding value being 
$\frac{32}{(3l+4)^3}\ (l+2)$. For \emph{odd}-$l$, the same occurs for $n=\frac{3}{2}(l+1)$, with a value 
$\frac{32}{27}\frac{1}{(l+1)^2}$. The bottom row (panels A(a)-A(b)) of Fig.~1 illustrates these variations of IR$_{\rvec}$ with 
changes in $n$ for lowest four \emph{even-} and \emph{odd-l} states respectively. Each curve passes 
through a maximum, which tends to shift towards right as $l$ assumes higher values. One also notices that when $n\gg l$, we 
achieve IR$_{\rvec} \approx -\frac{16}{Z^3}\mathcal{E}_{n}$.

\begin{figure}                         %%%Fig. 1, H-atom
\begin{minipage}[c]{0.45\textwidth}\centering
\includegraphics[scale=0.65]{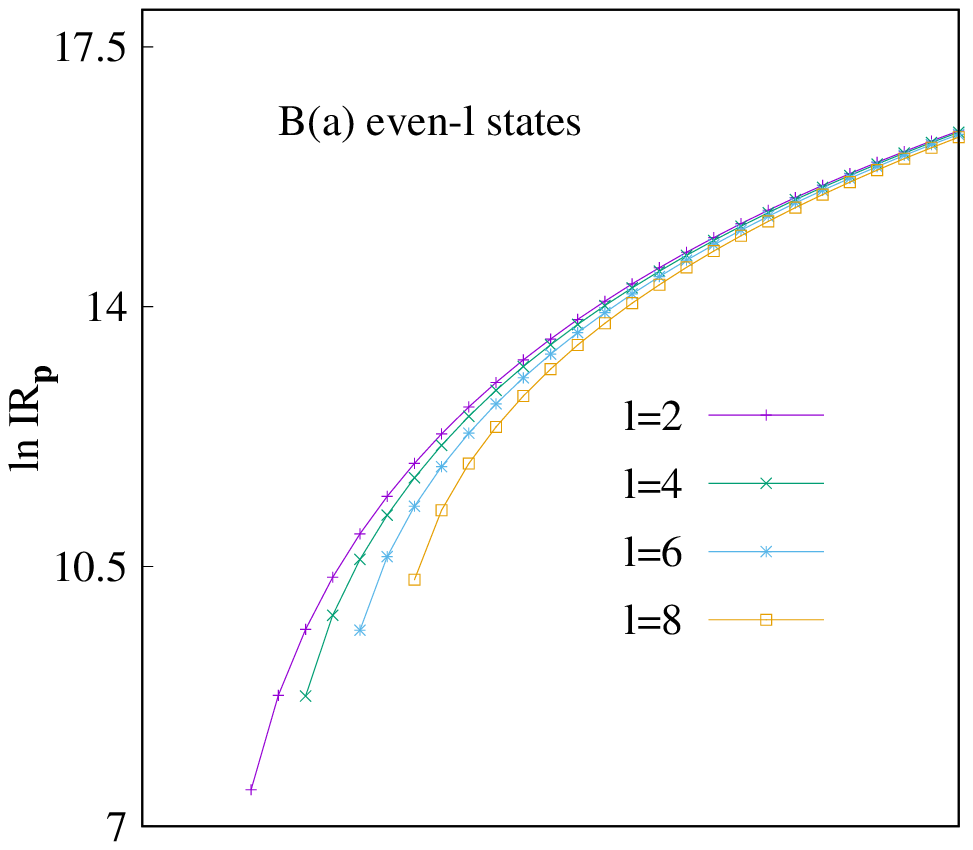}
\end{minipage}%
\begin{minipage}[c]{0.45\textwidth}\centering
\includegraphics[scale=0.65]{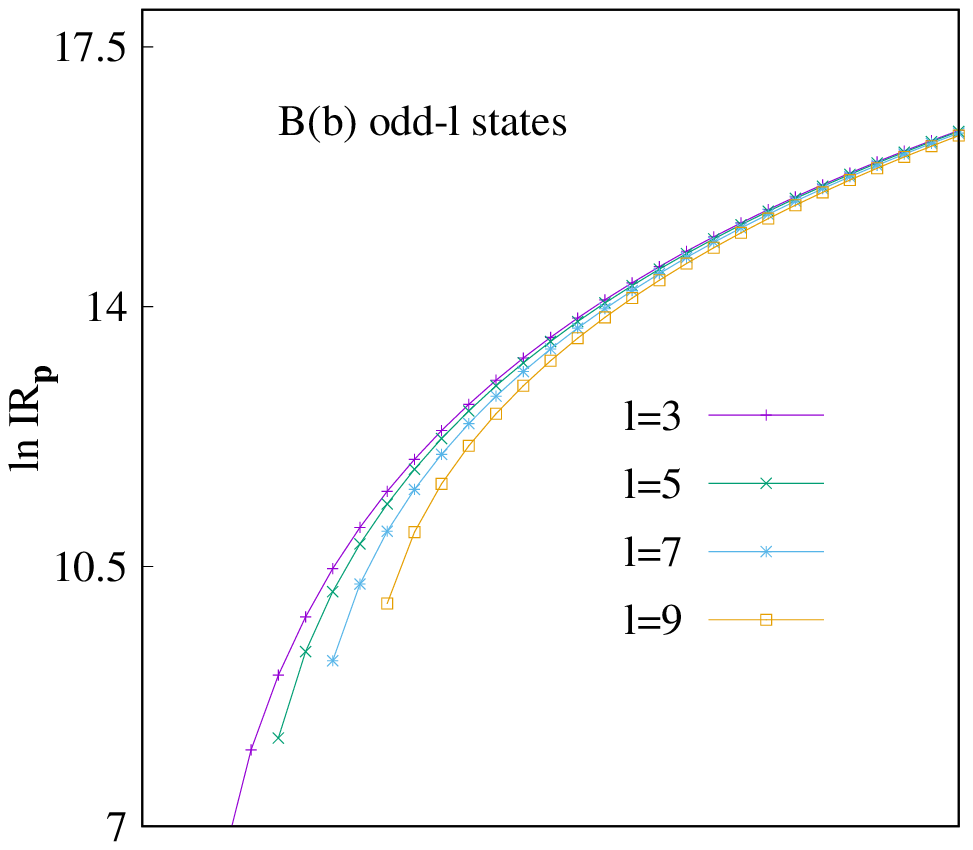}
\end{minipage}%
\vspace{0.1in}
\hspace{0.2in}
\begin{minipage}[c]{0.45\textwidth}\centering
\includegraphics[scale=0.7]{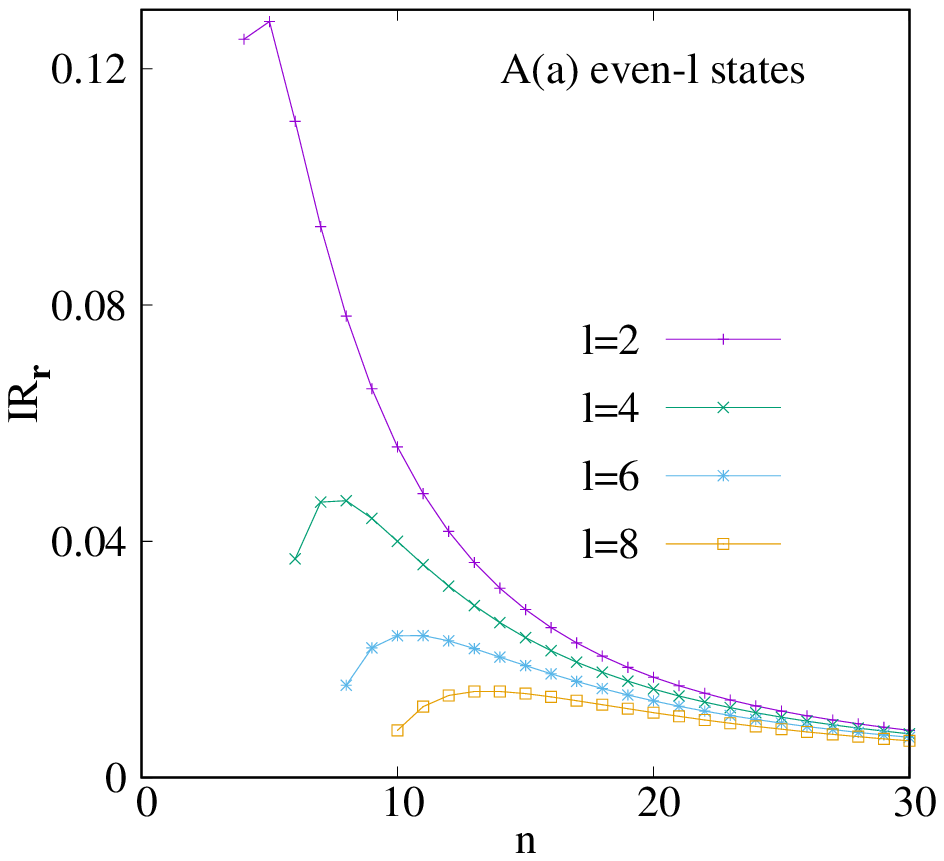}
\end{minipage}%
\begin{minipage}[c]{0.45\textwidth}\centering
\includegraphics[scale=0.7]{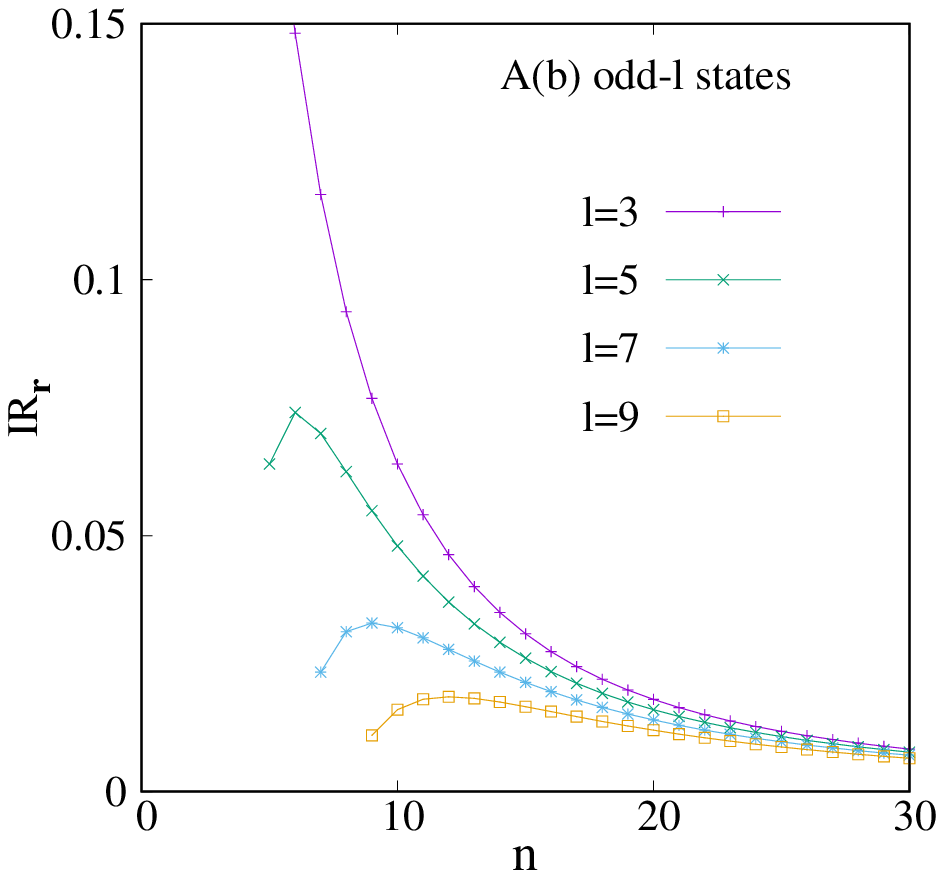}
\end{minipage}%
\caption{Variation of IR$_{\rvec}$ (A), IR$_{\pvec}$ (B) with $n$, for even-$l$ (a), odd-$l$ (b) states of H atom.}
\end{figure}

Next, we move on to IR$_{\pvec}$. The analytical expression \cite{sanudo08a} for wave function is given by,
\begin{equation}
\psi_{n,l}(p)=n^{2}\left[\frac{2}{\pi}\frac{(n-l-1)!}{(n+l)!}\right]^\frac{1}{2} 2^{(2l+2)} \ l! \ 
\frac{n^l}{ \{[\frac{np}{Z}]^2+1 \}^{l+2}} \left(\frac{p}{Z}\right)^l
C_{n-l-1}^{l+1} \left(\frac{[\frac{np}{Z}]^2-1}{[\frac{np}{Z}]^2+1}\right), 
\end{equation}
where $C_n^{\alpha}(x)$ signifies the Gegenbauer polynomial. Let us consider $t=\frac{np}{Z},$ and subsequently
$q=\frac{t^2-1}{t^2+1}$. These two substitutions transform Eq.~(20) into the form,
\begin{equation}
\mathrm{IR}_{\pvec} = Zl!^{2}n^{3}\left[\frac{2^{4l+7}}{\pi}\frac{(n-l-1)!}{(n+l)!}\right]
\int_{-1}^{1}(1+q)^{\frac{3}{2}}(1-q)^{\frac{5}{2}}[f_{n,l}(q)]^{2}
\left[\frac{d}{dq}\left(\frac{f_{n,l}^{\prime}(q)}{f_{n,l}(q)}-
\frac{f_{n_{1},l}^{\prime}(q)}{f_{n_{1},l}(q)}\right)\right]^{2}\mathrm{d}q,  
\end{equation}
where $f_{k,l}^{\prime}(q)=\frac{\mathrm{d}}{\mathrm{d}q}[f_{k,l}(q)]$ and 
$f_{(k,l)}(q)=(1+q)^{\frac{l}{2}}(1-q)^{\frac{l}{2}+2}C_{k-l-1}^{l+1}(q).$ Further we note that,
\begin{equation}
\begin{aligned}
\frac{f_{n,l}^{\prime}(q)}{f_{n,l}(q)} & =\frac{l}{2(1+q)}-\frac{\frac{l}{2}+2}{1-q}+
\frac{\left(C_{n-l-1}^{l+1}(q)\right)^{\prime}}{C_{n-l-1}^{l+1}(q)} \\
\frac{f_{n_{1},l}^{\prime}(q)}{f_{n_{1},l}(q)} & =\frac{l}{2(1+q)}-
\frac{\frac{l}{2}+2}{1-q}+\frac{\left(C_{n_{1}-l-1}^{l+1}(q)\right)^{\prime}}{C_{n_{1}-l-1}^{l+1}(q)}. 
\end{aligned}
\end{equation}
Once again, as in case of $L_n^{\alpha}(x)$ for IR$_{\rvec}$, $C_{n_{1}-l-1}^{l+1}(q)$, being a part of circular state, is a 
constant and hence $(C_{n_{1}-l-1}^{l+1}(q))^{\prime}=0$. Therefore, one may write, 
\begin{equation}
\frac{f_{n_{1},l}^{\prime}(q)}{f_{n_{1},l}(q)}=\frac{l}{2(1+q)}-\frac{\frac{l}{2}+2}{1-q}.
\end{equation}
After going through some algebra, one gets the following expression, 
\begin{equation}
\mathrm{IR}_{\pvec}=Z \ l!^{2} \ n^{3}\left[\frac{2^{4l+7}}{\pi}\ \frac{(n-l-1)!}{(n+l)!}\right](l+1)^{2}\left(I_{1}+I_{2}\right), 
\end{equation}
where the two integrations are defined as, 
\begin{equation}
I_{1}  =\int_{-1}^{1}(1-q^{2})^{l+\frac{3}{2}} \ \left[C_{n-l-2}^{l+2}(q)\right]^{2} \mathrm{d}q,  \ \ \ \ \ \ 
I_{2}  =\int_{-1}^{1}q(1-q^{2})^{l+\frac{3}{2}}\ \left[C_{n-l-2}^{l+2}(q)\right]^{2} \mathrm{d}q.
\end{equation}  
Using the fact that, $I_{2}=0$ as the integrand is an odd function of $q$, finally we obtain,  
\begin{equation}
\begin{aligned}
\mathrm{IR}_{\pvec} & =Z \ l!^{2} \ n^{3}\left[\frac{2^{4l+7}}{\pi}\frac{(n-l-1)!}{(n+l)!}\right](l+1)^{2}I_{1} \\
           & = 2^{4} \ Z \ n^{2}[n^{2}-(l+1)^{2}] \ \ \ (\mathrm{when} \ n>l, \ n-l \geq 2).
\end{aligned}
\end{equation}
This equation indicates that, at a constant $l$, IR$_{\pvec}$ enhances with $n$. On the contrary, at a fixed $n$, like IR$_{\rvec}$, 
it decreases with $l$. In panels B(a), B(b) of Fig.~1, $\mathrm{ln}(\mathrm{IR}_{\pvec})$ is plotted against $n$ for the same set 
of even, odd states as in IR$_{\rvec}$. These two graphs infer that, there is neither a maximum nor a 
minimum in IR$_{\pvec}$. In this occasion, when $n\gg l$, IR$_{\pvec} \approx \frac{4Z^{5}}{\mathcal{E}_{n}^{2}}$. 
No literature work exists for IR$_{\pvec}$ for direct comparison.

IR$_{\rvec}$ measures the fluctuation from lowest (reference) to a high-lying excited state (for a fixed $l$). Note that 
IR$_{\rvec}$, due to its inherent dependence on $l$, provides a more detailed information than Fisher information in H atom, 
because the latter has no such influence from $l$ \cite{romera05}. Also it reinforces the enhanced diffused nature of an orbital 
with $n$ for a fixed $l$. 

\subsubsection{Pseudoharmonic potential}
The $r$-space radial wave function for PHP is given as \cite{yahya15},
\begin{equation}
R_{n_{r},l}(r)=\left[\frac{2(2\lambda)^{\frac{(2\gamma_{l}+3)}{2}}n_{r}!}{\Gamma(n_{r}+\gamma_{l}+
\frac{3}{2})}\right]^{\frac{1}{2}}r^{\gamma_{l}}e^{-\lambda r^{2}} L_{n_{r}}^{\gamma_{l}+\frac{1}{2}}(2\lambda r^{2}), 
\end{equation}
where $\gamma_{l}=\frac{1}{2}\left[-1+\sqrt{(2l+1)^{2}+8\mu D_{e}r_{e}^{2}}\right]$, $\lambda=\sqrt{\frac{\mu D_{e}}{2r_{e}^{2}}}$ 
and $L_{n}^{\alpha}(x)$ refers to usual polynomial. Now replacing $\lambda r^{2}=\alpha$ and using the definition of IR given 
in Eq.~(20) we obtain,
\begin{equation}
\mathrm{IR}_{\rvec}=\left[\frac{2 \ (2\lambda)^{\frac{(2\gamma_{l}+3)}{2}} \ n_{r}!}{\Gamma(n_{r}+\gamma_{l}+\frac{3}{2})}\right] 
\frac{2}{(2\lambda)^{\gamma_{l}+1}} \int_{0}^{\infty}\alpha^{\gamma_{l}+\frac{3}{2}} \ e^{-\alpha} 
\left[L_{n_{r}}^{\gamma_{l}+\frac{1}{2}}(\alpha)\right]^{2} \left[\frac{R_{n_{r},l}^{\prime}(\alpha)}{R_{n_{r},l}(\alpha)}-
\frac{R_{n_{r_{1}},l}^{\prime}(\alpha)}{R_{n_{r_{1}},l}(\alpha)}\right]^{2} \mathrm{d} \alpha. 
\end{equation}  
Defining 
$R_{n_{r},l}(\alpha)=\left[\frac{2 \ (2\lambda)^{\frac{(2\gamma_{l}+3)}{2}} \ n_{r}!}{\Gamma(n_{r}+\gamma_{l}+
\frac{3}{2})}\right]^{\frac{1}{2}} \left(\frac{\alpha}{2\lambda}\right)^{\frac{\gamma_{l}}{2}}
e^{-\frac{\alpha}{2}}L_{n_{r}}^{\gamma_{l}+\frac{1}{2}}(\alpha)$, one gets, 
\begin{equation}
\frac{R_{n_{r},l}^{\prime}(\alpha)}{R_{n_{r},l}(\alpha)}=\frac{\gamma_{l}}{2\alpha}-1+\frac{\left\{L_{n_{r}}^{\gamma_{l}+
\frac{1}{2}}(\alpha) \right\}^{\prime}}{L_{n_{r}}^{\gamma_{l}+\frac{1}{2}}(\alpha)}, 
\end{equation}
where the prime denotes 1st derivative with respect to $\alpha$. For the reference state at a fixed $l$, $n_{r_{1}}=0$. Thus  
$L_{0}^{\gamma_{l}+\frac{1}{2}}(\alpha)=1$ and $[L_{n_{r_{1}}}^{\gamma_{l}+\frac{1}{2}}(\alpha)]^{\prime}=0$; hence 
one can write, 
\begin{equation}
\frac{R_{0,l}^{\prime}(\alpha)}{R_{0,l}(\alpha)}=\frac{\gamma_{l}}{2\alpha}-1. 
\end{equation} 
Finally invoking Eq.~(44), we find the following simplified expression, 
\begin{equation}
\begin{aligned}
\mathrm{IR}_{\rvec} &=\frac{8 \ (2\lambda)^{\frac{(2\gamma_{l}+3)}{2}} \ n_{r}!}{\Gamma(n_{r}+\gamma_{l}+\frac{3}{2})} \ 
\int_{0}^{\infty}\alpha^{\gamma_{l}+\frac{3}{2}} \ e^{-\alpha}\left[L_{n_{r}-1}^{\gamma_{l}+\frac{3}{2}}(\alpha)\right]^{2} 
\mathrm{d}\alpha  \\
           &=\frac{32n_{r}}{r_{e}}\sqrt{\frac{\mu D_{e}}{2}}.
\end{aligned}
\end{equation}

\begingroup           %%Table 2
\squeezetable
\begin{table}
\caption{Spectroscopic parameters for molecules considered, along with references.}
\begin{ruledtabular}
\begin{tabular}{c|c|c|c|c|c}
Molecule & State    & $\mu$ (amu)   & $D_{e}$ (eV)  & $r_{e}$ (\AA)    &   Reference     \\ 
\hline
H$_{2}$ &  X $^{1}\Sigma_{g}^{+}$   & 0.50391     & 4.7446       & 0.7416  &    \cite{oyewumi12}    \\
Na$_2$  &  X $^{1}\Sigma_{g}^{+}$   & 11.4948845  & 0.746707167  & 3.079   &    \cite{yahya15}    \\
Cl$_2$  &  X $^{1}\Sigma_{g}^{+}$   & 17.7275     & 2.513903386  & 1.987   &    \cite{yahya15}    \\
O$_{2}^{+}$ & X $^{2}\Pi_{g}$       & 7.9995      & 6.780447346  & 1.116   &    \cite{yahya15}    \\
CO      &  X $^{1}\Sigma^{+}$       & 6.860586000 & 10.845073641 & 1.1283  &    \cite{oyewumi12}    \\         
NO  &   X $^{2}\Sigma_{r}$          & 7.46844100  & 8.043729855  & 1.1508  &    \cite{oyewumi12}    \\
\end{tabular}
%%$^{\dag} Conversion factors: $1 eV = 0.03615384 a.u.~and~1$A^{0}$=1.88971616 a.u. \\
\end{ruledtabular}
\end{table}
\endgroup

Analogously we may proceed for the $p$-space IR, using respective wave function \cite{yahya15}, 
\begin{equation}
R_{n_{r},l}(p)=\left[\frac{2~n_{r}!}{(2\lambda)^{\frac{(2\gamma_{l}+3)}{2}}\Gamma(n_{r}+\gamma_{l}+
\frac{3}{2})}\right]^{\frac{1}{2}}p^{\gamma_{l}}e^{-\frac{p^{2}}{4\lambda}}
L_{n_{r}}^{\gamma_{l}+\frac{1}{2}}\left(\frac{p^{2}}{2\lambda}\right).
\end{equation}
Now putting $\beta=\frac{p^{2}}{2\lambda}$ and working out some standard algebra, we achieve, 
\begin{equation}
\mathrm{IR}_{\pvec}=\frac{2~n_{r}!}{(2\lambda)^{\frac{1}{2}}\Gamma(n_{r}+\gamma_{l}+\frac{3}{2})} \ 
\sqrt{\frac{2}{\lambda}}
\int_{0}^{\infty}\beta^{\gamma_{l}+\frac{3}{2}} \ e^{-\beta} \left[L_{n_{r}}^{\gamma_{l}+\frac{1}{2}}(\beta)\right]^{2}
\left[\frac{R_{n_{r},l}^{\prime}(\beta)}{R_{n_{r},l}(\beta)}-\frac{R_{n_{r_{1}},l}^{\prime}(\beta)}{R_{n_{r_{1}},l}(\beta)}
\right]^{2} \mathrm{d}\beta .
\end{equation}
Here we have defined, 
\begin{equation}
R_{n_{r},l}(\beta)=\left[\frac{2~n_{r}!}{(2\lambda)^{\frac{(2\gamma_{l}+3)}{2}}\Gamma(n_{r}+\gamma_{l}+\frac{3}{2})}
\right]^{\frac{1}{2}} \left(2\lambda \beta\right)^{\frac{\gamma_{l}}{2}}
e^{-\frac{\beta}{2}}L_{n_{r}}^{\gamma_{l}+\frac{1}{2}}(\beta). 
\end{equation}
Applying the same argument of reference state as before, leads to,
\begin{equation}
\frac{R_{n_{r},l}^{\prime}(\beta)}{R_{n_{r},l}(\beta)}-\frac{R_{n_{r_{1}},l}^{\prime}(\beta)}{R_{n_{r_{1}},l}(\beta)}
=- \ \frac{L_{n_{r}-1}^{\gamma_{l}+\frac{3}{2}}(\beta)}{L_{n_{r}}^{\gamma_{l}+\frac{1}{2}}(\beta)}
\end{equation} 
After some straightforward algebra, in the end, IR$_{\pvec}$ eventually takes the form,
\begin{equation}
\begin{aligned}
\mathrm{IR}_{\pvec} &=4\left[\frac{2~n_{r}!}{(2\lambda)^{\frac{1}{2}}\Gamma(n_{r}+\gamma_{l}+\frac{3}{2})}\right] 
\sqrt{\frac{2}{\lambda}}
\int_{0}^{\infty}\beta^{\gamma_{l}+\frac{3}{2}}e^{-\beta}\left[L_{n_{r}-1}^{\gamma_{l}+\frac{3}{2}}(\beta)\right]^{2} 
\mathrm{d}\beta  \\
           &=8n_{r}r_{e}\sqrt{\frac{2}{\mu D_{e}}}.
\end{aligned}
\end{equation}

\begingroup           %%Table 3
\squeezetable
\begin{table}
\caption{IR$_{\rvec}$, IR$_{\pvec}$ for H$_{2}$, Na$_{2}$, Cl$_{2}$, O$_{2}^{+}$, CO, NO at selected $n_r$. 
All results in atomic unit.}
\centering
\begin{ruledtabular}
\begin{tabular}{c|cc|cc|cc}
 $n_r$   & \multicolumn{2}{c|}{H$_{2}$}  & \multicolumn{2}{c|}{Na$_{2}$} &  \multicolumn{2}{c}{Cl$_{2}$}  \\
\cline{2-3} \cline{4-5}  \cline{6-7} 
                    & IR$_{\rvec}$ & IR$_{\pvec}$   & IR$_{\rvec}$ & IR$_{\pvec}$  & IR$_{\rvec}$ & IR$_{\pvec}$   \\
\hline
                        1   & 202.676044   & 1.263099   & 92.494014    & 2.767746   & 326.584840   & 0.783869 \\
                        2   & 405.352089   & 2.526198   & 184.988028   & 5.535493   & 653.169681   & 1.567739 \\
                        3   & 608.028133   & 3.789298   & 277.482042   & 8.303240   & 979.754522   & 2.351609 \\
                       10   & 2026.760446  & 12.630994  & 924.940140   & 27.677466  & 3265.848407  & 7.838698 \\
                       25   & 5066.901116  & 31.577486  & 2312.350352  & 69.193666  & 8164.621018  & 19.596745 \\
                       50   & 10133.802233 & 63.154972  & 4624.700704  & 138.387333 & 16329.242036 & 39.193490 \\ 
                      100   & 20267.604466 & 126.309944 & 9249.401408  & 276.774667 & 32658.484073 & 78.386981 \\
\hline
        & \multicolumn{2}{c|}{O$_{2}^{+}$}  & \multicolumn{2}{c|}{CO}  & \multicolumn{2}{c}{NO}  \\
\hline
                        1   & 641.493486   & 0.399068  & 743.135693   & 0.344486   & 654.696038   & 0.391021 \\
                        2   & 1282.986973  & 0.798137  & 1486.271387  & 0.688972   & 1309.392076  & 0.782042 \\
                        3   & 1924.480460  & 1.197206  & 2229.407081  & 1.033458   & 1964.088114  & 1.173063 \\
                        10  & 6414.934868  & 3.990687  & 7431.356937  & 3.444862   & 6546.960382  & 3.910211 \\
                        25  & 16037.337170 & 9.976718  & 18578.392343 & 8.612155   & 16367.400957 & 9.775528 \\
                        50  & 32074.674340 & 19.953437 & 37156.784686 & 17.224310  & 32734.801914 & 19.551057 \\
                       100  & 64149.348680 & 39.906874 & 74313.569373 & 34.448621  & 65469.603828 & 39.102115 \\
\end{tabular}
\end{ruledtabular}
\end{table}
\endgroup

Equations~(47) and (52) suggest that, both IR$_{\rvec}$, IR$_{\pvec}$ change linearly with $n_{r}$. Like the case of QHO in 1D and
3D, in this occasion also, IR in both spaces remain invariant for any two successive states as
given by the equation below, 
\begin{equation}
\begin{aligned}
\Delta \mathrm{IR}_{\rvec} & = \mathrm{IR}_{\rvec}(n_{r}+1)-\mathrm{IR}_{\rvec}(n_{r})=\frac{32}{r_{e}}\sqrt{\frac{\mu D_{e}}{2}} \\
\Delta \mathrm{IR}_{\pvec} & = \mathrm{IR}_{\pvec}(n_{r}+1)-\mathrm{IR}_{\pvec}(n_{r})= 8r_{e} \sqrt{\frac{2}{\mu D_{e}}}.  
\end{aligned}
\end{equation} 
In order to provide some numerical data, we have selected six homo- and hetero-nuclear diatomic molecules, namely, H$_2$, Na$_2$, 
Cl$_2$, O$_{2}^{+}$, CO and NO, including a cation. Using the parameters, $\mu, D_{e}, r_{e}$ listed in Table~II, as quoted 
from \cite{yahya15, oyewumi12}, adopting following conversion factors, 1 amu = 1.82289$\times 10^3$ a.u., 
1 eV = 0.03615384 a.u., 1 \AA =1.88971616 a.u., and exploiting Eqs.~(47), (52) we have computed IR$_{\rvec}$, IR$_{\pvec}$ for all 
these species. These are tabulated in Table~III for 7 selective $n_r$, \emph{viz.}, 1,~2,~3,~10,~25,~50, 100 respectively.  
One notices that like 3D QHO, in PHP case also, IR$_{\rvec}$ complements the findings of Fisher information reported elsewhere 
\cite{romera05, yahya15}. Moreover, for both these potentials, a growth in $n_{r}$ causes raising of relative fluctuation, which 
consequently results in a weakening of bond strength. 

\section{Future and Outlook}
In this work, we have derived generalized expressions for IR for an arbitrary quantum state in a 1D QHO, as well as three 
central potentials, \emph{viz.}, 3D QHO, H atom and PHP. In the former the ground state is considered as reference, while for 
latter, the lowest state corresponding to a given $l$ was employed for same. In 1D QHO, IR in both spaces 
vary linearly with state index $n$. However the variation with respect to $w$ contrasts each other in two spaces; in $x$ space, 
it has linear dependence while in $p$ space, it is inversely proportional. Interestingly the difference in IR  
in adjacent states remains constant in both conjugate spaces, for a given $\omega$. In H atom, IR$_{\rvec}$ lowers while 
IR$_{\pvec}$ grows with $n$; but both reduce as $l$ advances. In contrary, in 3D QHO and PHP, both measures progress 
linearly with $n_{r}$. In 3D QHO, IR's show opposite trends in conjugate spaces with changes in $\omega$. Further, behavioral
pattern of these with respect to $Z$ is also considered in H atom. A detailed inspection of IR as well as other 
relative information measure like R\'enyi and Shannon entropy, Onicescu energy in the context of 
other potentials of physical/chemical interest including atoms, molecules may be worthwhile and desirable. 
Most of the results presented here are new. 

\section{Acknowledgement}
Financial support from DST SERB, New Delhi, India (sanction order: EMR/2014/000838) is gratefully acknowledged. NM thanks DST SERB, 
New Delhi, India, for a National-post-doctoral fellowship (sanction order: PDF/2016/000014/CS).

\end{document}